%% file: 0PBH_header.tex
\documentclass[onecolumn]{aastex6}
\pdfoutput=1
\usepackage{xcolor}                                 

\usepackage{amsmath}
\usepackage{amsfonts}
\usepackage{dsfont}
\usepackage{amsxtra}
\usepackage{hyperref}
\usepackage{amssymb}
\usepackage{upgreek}
\usepackage{multirow}
\usepackage{url}

\usepackage{graphicx,epsfig}
\usepackage{dcolumn}
\usepackage{bm}

\usepackage{xspace} 

\usepackage{lineno}

\newcommand{\be}{\begin{equation}}
\newcommand{\ee}{\end{equation}}
\newcommand{\bea}{\begin{eqnarray}}
\newcommand{\eea}{\end{eqnarray}}
\newcommand{\beaa}{\begin{eqnarray*}}
\newcommand{\eeaa}{\end{eqnarray*}}
\newcommand{\ba}{\begin{array}}
\newcommand{\ea}{\end{array}}
\newcommand{\bi}{\begin{itemize}}
\newcommand{\ei}{\end{itemize}}
\newcommand{\ben}{\begin{enumerate}}
\newcommand{\een}{\end{enumerate}}

\newcommand{\adeg}{^\circ\!\!}

\newcommand{\lb}{\label}
\newcommand{\g}{\gamma}

\newcommand{\Om}{\Omega}

\newcommand{\La}{{\mathcal{L}}}

\newcommand{\Fermi}{\textsl{Fermi}\xspace}

\newcommand{\blue}{}
\definecolor{darkgreen}{rgb}{0.0, 0.7, 0.0}

\newcommand{\onepic}{.4}
\newcommand{\twopic}{.4}

\newcommand{\rbh}{\dot{\rho}_{\rm pbh}}

\newcommand{\Ombh}{\Om_{\rm pbh}}

\newcommand{\rhoc}{{\rho_{\rm c}}}

\shorttitle{Search for Gamma Ray Emission from Local Primordial Black Holes with the \Fermi Large Area Telescope}
\setwatermarkfontsize{1.5in}

\begin{document}

\title{SEARCH FOR GAMMA-RAY EMISSION FROM LOCAL PRIMORDIAL BLACK HOLES WITH THE {\it FERMI}
LARGE AREA TELESCOPE}

\author{
M.~Ackermann\altaffilmark{1}, 
W.~B.~Atwood\altaffilmark{2}, 
L.~Baldini\altaffilmark{3}, 
J.~Ballet\altaffilmark{4}, 
G.~Barbiellini\altaffilmark{5,6}, 
D.~Bastieri\altaffilmark{7,8}, 
R.~Bellazzini\altaffilmark{9}, 
B.~Berenji\altaffilmark{10}, 
E.~Bissaldi\altaffilmark{11,12}, 
R.~D.~Blandford\altaffilmark{13}
% ...
}
\altaffiltext{1}{Deutsches Elektronen Synchrotron DESY, D-15738 Zeuthen, Germany}
\altaffiltext{2}{Santa Cruz Institute for Particle Physics, Department of Physics and Department of Astronomy and Astrophysics, University of California at Santa Cruz, Santa Cruz, CA 95064, USA}
\altaffiltext{3}{Universit\`a di Pisa and Istituto Nazionale di Fisica Nucleare, Sezione di Pisa I-56127 Pisa, Italy}
\altaffiltext{4}{Laboratoire AIM, CEA-IRFU/CNRS/Universit\'e Paris Diderot, Service d'Astrophysique, CEA Saclay, F-91191 Gif sur Yvette, France}
\altaffiltext{5}{Istituto Nazionale di Fisica Nucleare, Sezione di Trieste, I-34127 Trieste, Italy}
\altaffiltext{6}{Dipartimento di Fisica, Universit\`a di Trieste, I-34127 Trieste, Italy}
\altaffiltext{7}{Istituto Nazionale di Fisica Nucleare, Sezione di Padova, I-35131 Padova, Italy}
\altaffiltext{8}{Dipartimento di Fisica e Astronomia ``G. Galilei'', Universit\`a di Padova, I-35131 Padova, Italy}
\altaffiltext{9}{Istituto Nazionale di Fisica Nucleare, Sezione di Pisa, I-56127 Pisa, Italy}
\altaffiltext{10}{California State University, Los Angeles, Department of Physics and Astronomy, Los Angeles, CA 90032, USA}
\altaffiltext{11}{Dipartimento di Fisica ``M. Merlin" dell'Universit\`a e del Politecnico di Bari, I-70126 Bari, Italy}
\altaffiltext{12}{Istituto Nazionale di Fisica Nucleare, Sezione di Bari, I-70126 Bari, Italy}
\altaffiltext{13}{W. W. Hansen Experimental Physics Laboratory, Kavli Institute for Particle Astrophysics and Cosmology, Department of Physics and SLAC National Accelerator Laboratory, Stanford University, Stanford, CA 94305, USA}
% ...

\email{Corresponding authors: Christian Johnson, Stefan Funk, Dmitry Malyshev, Steven Ritz}

\begin{abstract}
Black holes with masses below approximately $10^{15}$ g are expected to emit gamma rays with energies above a few tens of MeV, which can be detected by the \Fermi Large Area Telescope (LAT).
Although black holes with these masses cannot be formed as a result of stellar evolution, they may have formed in the early Universe and are therefore called Primordial Black Holes (PBHs).
Previous searches for PBHs have focused on either short timescale bursts or the contribution of PBHs to the isotropic gamma-ray emission.
We show that, in case of individual PBHs, the \Fermi LAT is most sensitive to PBHs with temperatures above approximately 16 GeV 
and masses $6\times 10^{11}$ g, which it can detect out to a distance of about 0.03 pc.
These PBHs have a remaining lifetime of months to years at the start of the \Fermi mission.
They would appear as potentially moving point sources with gamma-ray emission that becomes spectrally harder and brighter with time until the PBH completely evaporates.
In this paper, we develop a new algorithm to detect the proper motion of a gamma-ray point sources, and apply it to 318 unassociated point sources at high galactic latitude in the third \Fermi-LAT source catalog (3FGL).
None of unassociated point sources with spectra consistent with PBH evaporation show significant proper motion.
Using the non-detection of PBH candidates, we derive a 99\% confidence limit on PBH evaporation rate in the vicinity of the Earth 
$\dot{\rho}_{\rm PBH} < 7.2 \times 10^3\: {\rm {pc}^{-3} {yr}^{-1}}$.
This limit is similar to the limits obtained with ground-based gamma-ray observatories.

\end{abstract}




\input{1PBH_intro} 
\input{2PBH_volume} 
\newpage
\input{3PBH_PSstudy} 
\input{4PBH_MClimits} 
\input{5PBH_discussion} 
\input{acknowledgement}

\appendix
\input{App_PBH_spectra} 

\bibliography{BHpapers}

\end{document}

%% file: 1PBH_intro.tex
\section{Introduction}
\lb{sec:intro}

The formation of primordial black holes (PBHs)
is a prediction of some models of the early Universe
\citep{1966AZh....43..758Z, 1971MNRAS.152...75H}.  
In this paper, we search for evidence of gamma rays produced by the Hawking radiation of low-mass, high-temperature PBHs in the \Fermi Large Area Telescope (LAT) data.

In classical general relativity, \blue{black holes have zero temperature and do not emit particles.}
The concept of a black hole temperature was formally introduced
as a way to resolve the paradox of information loss for matter falling
into the black hole \citep{1973PhRvD...7.2333B, 1974PhRvD...9.3292B}.
\cite{1974Natur.248...30H, 1975CMaPh..43..199H} argued that black holes should emit particles
due to pair creation near the horizon with grey body emission spectra with the temperature

\noindent
\be
\label{eq:bh_mass}
T_{\rm BH} = \frac{\hbar c^3}{8 \pi G M k} \approx 10^{-7} 
\left( \frac{M}{M_\odot} \right)^{-1} K,
\ee 
where $M$ is the black hole mass and $M_\odot$ is the Solar mass.
Astrophysical black holes, such as those created in a collapse of a massive star,
have masses larger than $M_\odot$. 
The corresponding temperature is less than $10^{-7} $ K, 
which is less than the fluctuations in the cosmic microwave background (CMB) radiation; 
this makes the signal from these BHs extremely difficult to detect.
However, PBHs with masses as small as the Planck mass ($\sim 10^{-5}$ g) may have been created 
in the early Universe \citep{1966AZh....43..758Z, 1971MNRAS.152...75H}.


The mass of a PBH created at time $t$ after the Big Bang is
proportional to the mass within the particle horizon at time $t$ 
\citep[for a review see][]%
{1991Natur.353..807H,2005astro.ph.11743C,2010RAA....10..495K},

\noindent
\be 
M(t) \approx10^{15} \left( \frac{t}{10^{-23}{\rm s}} \right) {\rm g}.  
\ee
Limits on density of PBHs at mass $M$ constrain the magnitude of density
fluctuations (or the equation of state) at time $t$.
These constraints are complementary to the constraints obtained from
observations of the CMB since they are sensitive to fluctuations on spatial scales much smaller than the scales observed in the CMB
\citep[e.g.,][]{2013PhRvD..87j3506L}.

Since Hawking radiation is emitted in all available particle species, the total emitted power and therefore the lifetime of a PBH depends on the number of available particle states. For the Standard Model of particles, the lifetime of a PBH can be approximated as
\citep{2010PhRvD..81j4019C} 

\noindent
\be 
\tau \approx 400 \left( \frac{M}{10^{10} {\rm g}} \right)^3 {\rm s}.  
\ee 
The PBHs which were formed in the early Universe with a mass around $M_* = 5\times 10^{14}$ g
have a lifetime close to the lifetime of the Universe and reach late stages of evaporation at the present time.
The PBHs with the remaining lifetime of months to years will be the subject of the search in this paper.
The temperature corresponding to $M_*$ is $T_{\rm BH} \approx 20$ MeV \citep{1991PhRvD..44..376M, 2010PhRvD..81j4019C}.%
\footnote{Here and in the following we convert temperature to energy units
by multiplying with the Boltzmann constant $k$.
}
A limit on the average cosmological density of PBHs can be obtained by integrating the flux from PBHs over the lifetime of the Universe.
This flux has to be smaller than the observed extragalactic gamma-ray background \citep{1998PhR...307..141C, 2010PhRvD..81j4019C}.
This method is most sensitive to PBHs with initial masses $M \approx M_*$.
Since PBHs are expected to be concentrated in galaxies similarly to dark matter,
stronger constraints for initial masses slightly larger than
$M_*$ can be obtained by searching for diffuse emission from PBHs in the halo of the Milky way galaxy
\citep{2009A&A...502...37L}.
A comparison of extragalactic and Galactic gamma-ray background constraints with the constraints from the search for individual PBHs will be presented in Section \ref{sec:discuss}.

Constraints on the local PBH evaporation rate have been obtained by looking for bursts of high-energy gamma-ray emission with
duration of a fraction of a second to several seconds. In particular, PBHs with a mass of \mbox{$M \sim 10^9$ g} have temperature 
$T_{\rm  BH} \approx (10^{10}\, {\rm g} / M_{\rm BH})\,{\rm TeV} \sim 10\,{\rm TeV}$ \citep{2010PhRvD..81j4019C}
and lifetime $\tau \sim 0.4$ sec. 
A search for such PBHs in the context of the Standard Model has been carried out by searching for high-energy gamma rays ($\approx$ 100 GeV $<$ E $< \approx$ 50 TeV) with Cherenkov telescopes
\citep[e.g.,][]{2012JPhCS.375e2024T, 2013arXiv1307.4898G, 2015APh....64....4A}.
In the case of the Hagedorn model \citep{1968NCimA..56.1027H},
\blue{which was proposed before the Standard Model of particle physics was confirmed,}
the number of states available for the PBH evaporation increases exponentially
when the temperature reaches the Hagedorn transition energy around 160 MeV. 
This leads to a microsecond burst of gamma rays around a few hundred MeV \citep{1976ApJ...206....1P}.
A search for such bursts was carried out using EGRET data \citep{1994ApJ...434..557F}. 
One can also expect short radio bursts in this model \citep{1977Natur.266..333R, 1977MNRAS.181..489B}, 
which can be used to constrain the PBH evaporation rate \cite[e.g.,][]{2015PASP..127.1269C}.

\Fermi LAT is a pair conversion telescope which is sensitive to gamma rays from $\approx$ 20 MeV to more than 300 GeV. Overviews of the \Fermi-LAT design and performance can be found in \citet{2009ApJ...697.1071A} and \citet{2012ApJS..203....4A}. 
PBHs with masses \mbox{$M \lesssim 10^{15}$ g} have temperatures \mbox{$T \gtrsim 10$ MeV}
and emit gamma rays \citep{1976ApJ...206....1P} that can be detected with the \Fermi LAT.


In this paper, we search for PBHs in the context of the Standard Model of particles using the \Fermi-LAT gamma-ray data. 
In Section \ref{sec:sens} we estimate that for the differential point source (PS) sensitivity in 4 years of observation
the \Fermi LAT is most sensitive to PBHs with temperature $T_{\rm  BH} \sim 16$ GeV(corresponding to a remaining lifetime of $\sim$4 years).
The corresponding distance to which such a PBH can be detected is $\lesssim 0.02$ pc. Assuming a random relative velocity of PBHs around the Earth similar to that of particle dark matter velocity dispersion,
the displacement of a PBH over 4 years within 0.02 pc is, on average, greater than $1^\circ$.
For comparison, the point source (PS) localization radius for a near-threshold source is approximately 0.1$^\circ$ \citep{2012ApJS..203....4A}.
Thus a smoking-gun signature of a PBH in \Fermi-LAT data is a moving source with a hard spectrum of gamma rays.
A search for such sources is a novel feature of our analysis that has not been discussed in the literature
related to PBH searches.

In Section \ref{sec:search} we search for PBH candidates among the sources of the third \Fermi-LAT catalog \citep[3FGL, ][]{2015ApJS..218...23A}.
The majority of sources can be excluded as PBH candidates based on association with known sources
or on their spectra, but several sources cannot be excluded based on these criteria alone. For these sources, we analyze the 
\Fermi-LAT data near the position of the PBH candidates and find one candidate whose proper motion is inconsistent with zero. 
We discuss this source in more detail in Section \ref{sec:j2310}.
In Section \ref{sec:MClimit}  
we use Monte Carlo simulations to derive the efficiency of our PBH selection criteria, 
which we then use to derive an upper limit on the PBH evaporation rate.
We find that PBHs can be detected up to a distance of $\sim$ 0.03 pc at the expense of a reduction of volume in phase space:
the PBHs should be moving preferentially along the line of sight to be detected as a PS.
The detectability distance of $\approx$0.03 pc derived with MC simulations and the full PS analysis pipeline is generally consistent with 
the simple estimate of 0.02 pc, which we obtain for non-moving PBHs using the known differential sensitivity of \Fermi LAT to point sources.
In Section \ref{sec:discuss} we present the conclusions.
In Appendix \ref{app:PBH_spectrum} we discuss some uncertainties on the spectrum of gamma rays emitted by PBHs.

%% file: 2PBH_volume.tex
\section{Sensitivity of \Fermi LAT to Individual Primordial Black Holes}
\lb{sec:sens}

We start our analysis by estimating the sensitivity domain, i.e., the relevant energies and timescales,  of the \Fermi LAT to individual PBHs.
One of the main questions is whether the \Fermi LAT is more sensitive to a population of PBHs with ``low" temperature (e.g., $\lesssim 10$ GeV), 
hence, smaller intensity of emission but large spatial density, or to a population with ``high" temperature (e.g., $\gtrsim 10$ GeV) 
but low spatial density.

\begin{figure}[htbp] 
\begin{center}
\epsfig{figure = 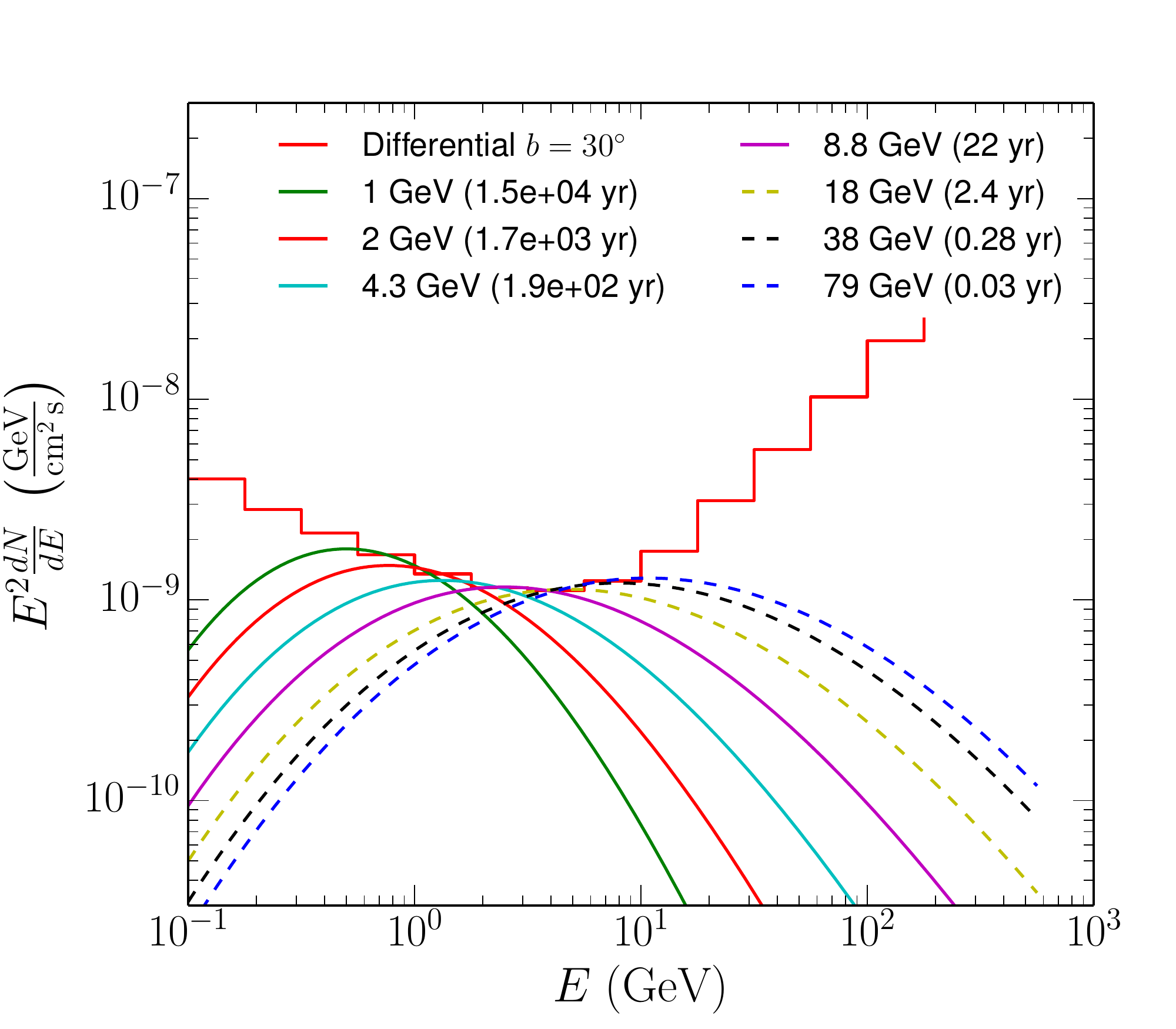, scale=\onepic}
\noindent
\caption{\small 
\label{fig:PSsensitivity}
Comparison of spectra of PBHs with different initial temperatures and \Fermi-LAT PS differential sensitivity at $b = 30^\circ$,
which is representative for \Fermi-LAT PS sensitivity away from the Galactic plane.
Lines correspond to PBHs with different initial temperatures specified in the labels, with the corresponding lifetimes shown in parentheses.
PBHs with lifetimes longer than 4 years are shown as solid lines, while PBHs with lifetime shorter than 4 years are shown as dashed lines.
The distance to each PBH is chosen such that the flux from the PBH is equal to the PS sensitivity in one of the energy bins
(the corresponding detectability distances are shown as a solid line in Figure \ref{fig:Rsens} on the right).
}
\end{center}
\end{figure}

In this section, we derive the range of masses, temperatures, and distance to the Earth where PBHs are detectable by \Fermi LAT.
In Figure \ref{fig:PSsensitivity}, we compare the spectra of PBHs with the differential PS sensitivity 
for 4 years of Pass 7 reprocessed data 
\citep{2012ApJS..203....4A, 2013arXiv1304.5456B}.
The same data sample was used in the derivation of the 3FGL catalog
\citep{2015ApJS..218...23A}, which we use in the next section to search for PBH candidates.
To determine the four-year equivalent flux from a PBH, we integrate the PBH spectrum either over the lifetime of the PBH or over four years, whichever is smaller, and divide by four years.
The PBH spectra are derived taking into account both primary and secondary (mostly from hadronic showers of quarks and gluons)
production of gamma rays by the black holes \citep{1991ApJ...371..447M}.
Although all Standard Model particles can be emitted by PBHs in principle, the $\gamma$-ray spectrum is dominated by the 
Quantum chromodynamics (QCD) degrees of freedom \citep{MacGibbon2015mya}.
The normalization for each of the curves is chosen such that the PBH flux is equal to the
differential PS sensitivity in one of the energy bins.

\begin{figure}[htbp] 
\begin{center}
\epsfig{figure = 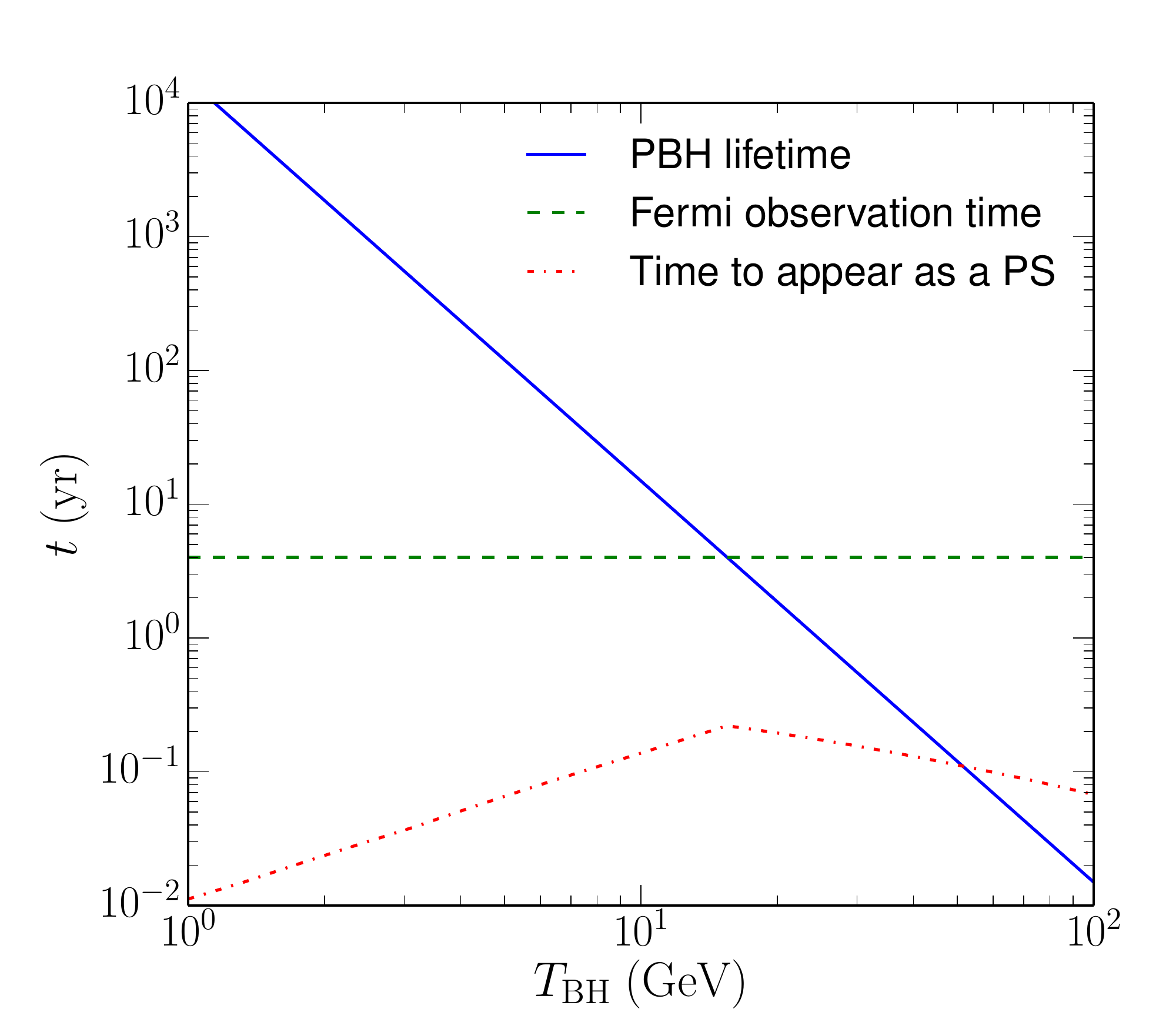, scale=\twopic}
\epsfig{figure = 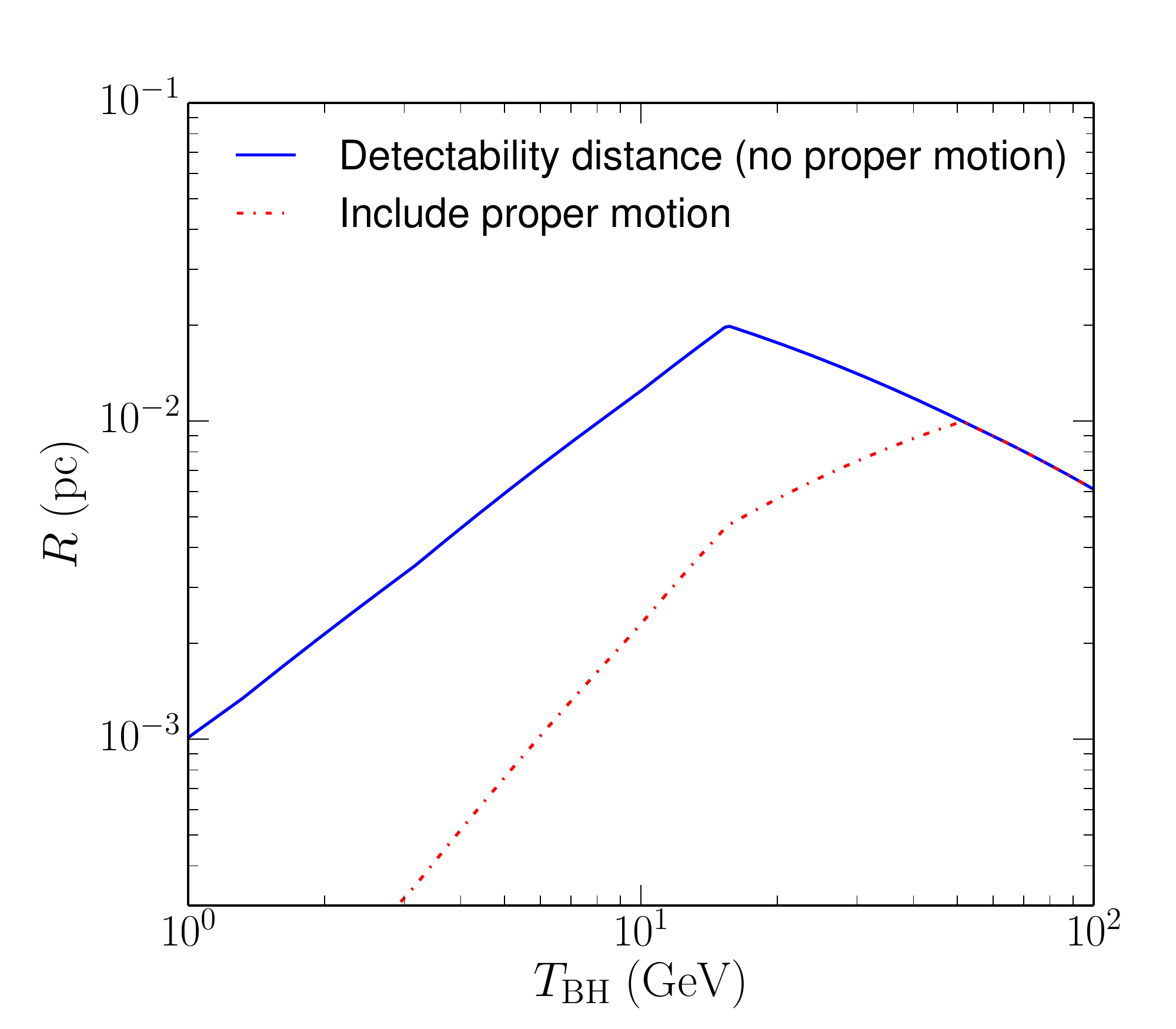, scale=\twopic}
\noindent
\caption{\small 
\label{fig:Rsens}
Left: solid line - PBH lifetime as a function of initial temperature, dashed line - 4 years, dash-dotted line - time during which the displacement is less than $0.2^\circ$ which is approximately equal to twice the PS localization radius at 10 GeV for Pass 7 reprocessed data assuming relative velocity perpendicular to the line of sight $\sim 300 \; {\rm km\: s^{-1}}$ and distance represented by the solid line on the right plot.
Right: solid line - detectability distance for a PBH as a function of initial temperature $T$ from the \Fermi-LAT PS differential sensitivity
(Figure \ref{fig:PSsensitivity}); dash-dotted line - detectability distance taking into account relative motion of PBHs due to 
orbital motion in the Galaxy and dark matter-like velocity dispersion (dash dotted line on the left plot).
PBHs with lifetimes of 4 years have a temperature of about 16 GeV; the break at 16 GeV results from lower-temperature PBHs having longer lifetimes than the observation time, i.e., they evaporate only partially, while PBHs with temperatures higher than 16 GeV have a 
smaller initial mass, i.e., produce smaller total fluxes than PBHs with initial temperature of 16 GeV.
}
\end{center}
\end{figure}

Since the intrinsic luminosity of PBHs for a given temperature is fixed, we can use the condition 
that the flux should be larger than the PS sensitivity
to estimate the maximal distance at which the PBHs can be detected as a function of initial temperature
(shown by the solid line in Figure \ref{fig:Rsens} right).
Typical values near the maximum are $\lesssim$ 0.02 pc for initial temperatures between 10 GeV and 40 GeV.
The corresponding remaining lifetime is between ~ 20 years and ~ 4 months.
Over the age of the Universe, we expect PBHs to virialize in the same manner as dark matter.
If we take into account the orbital velocity of the solar system around the GC $v_{\rm rot} \sim 250\; {\rm km\: s^{-1}}$
\citep{1999MNRAS.310..645W, 2008ApJ...684.1143X, 2009PASJ...61..227S, 2010ApJ...720L.108G, 
2010MNRAS.402..934M, 2011MNRAS.414.2446M}
and add a velocity dispersion similar to the expected dark matter velocity dispersion near the Sun $v_{\rm disp} \sim 270\; {\rm km\: s^{-1}}$
\citep[e.g.,][]{2010JCAP...02..030K}
then the displacement of a PBH during 3 years at $R = 0.02$ pc from the Earth could be as large as $\alpha = v_{\rm perp} t / R \propto 3^\circ$.
Here we estimated the average velocity perpendicular to the line of sight in a random direction on the sky as
$v_{\rm perp} = \sqrt{2/3 ( v_{\rm rot}^2 + v_{\rm disp}^2)} \sim 300\; {\rm km\: s^{-1}}$.
Because this displacement is $\sim$ 15 times larger than the \Fermi-LAT PS localization radius for a threshold source, such a black hole will appear as a linear streak of gamma-ray emission instead of a point source.


Since the large majority of \Fermi-LAT sources are point-like, linear spatial extension is a powerful criterium for identifying PBHs.
One of the difficulties in observing a moving PS is that the flux becomes smeared out into an extended track following the trajectory of the source, but PS catalogs are optimized for sensitivity to point sources as opposed to extended sources.
A simple estimate of the \Fermi-LAT sensitivity to a moving source can be obtained by
integrating the flux during the time when the moving source appears
as a point-like source, e.g., when the displacement is less than the PS localization radius.
The corresponding time is shown by in Figure \ref{fig:Rsens}, left.%
\footnote{The time for a PBH with an average velocity to appear as a point-like source is smaller than 4 years, which is the observation time
for the 3FGL catalog.
Better sensitivity to PBHs can be obtained by searching for PS with duration in the range from approximately a month to a year
due to lower background for shorter integration time.
Such a search of PS with a sliding time window goes beyond the scope of this work, where we restrict the analysis
to PS detected in the 3FGL catalog.}
Using the reduced integration time, we also derive the corresponding detectability distance 
(red dash-dotted line in Figure \ref{fig:Rsens} on the right). 
For large temperature (small remaining lifetime) the detectability distance for a moving source is the same as for a stationary one because the expected displacement is smaller than the PS localization radius.

Given the characteristic detection radius for PBHs including the proper motion $R \sim$ 0.01 pc 
for 4 years of observations (Figure \ref{fig:Rsens}, right panel),
we estimate that the \Fermi LAT is sensitive to a PBH evaporation rate of 
$\dot{\rho} = 1 / V t \propto 6\times 10^4\, {\rm pc^{-3} yr^{-1}}$, where $V = 4 \pi R^3 / 3$ is the detectability volume and $t = 4$ yr is the observations time.
The main result of this section is that the \Fermi-LAT sensitivity is potentially competitive with 
sensitivity to individual PBHs of Cherenkov observatories 
\citep[][]{2012JPhCS.375e2024T, 2013arXiv1307.4898G, 2015APh....64....4A, 2016APh....80...90U}, e.g., 
$\dot{\rho}  = 1.4 \times 10^4\, {\rm pc^{-3} yr^{-1}} $ derived by the H.E.S.S. collaboration
\citep{2013arXiv1307.4898G},
but one has to take the proper motion of PBHs into account.
In the following sections we use the 3FGL catalog, which employs the full PS sensitivity of the \Fermi LAT rather than the differential sensitivity.
We also treat the proper motion more rigorously.
As a result, the derived \Fermi-LAT sensitivity is a factor of a few better than the simple estimate presented in this section.

%% file: 3PBH_PSstudy.tex
\section{A Search for PBH Candidates in \Fermi-LAT 3FGL Catalog}
\label{sec:search}
The \Fermi-LAT surveys the entire sky approximately every three hours, and has relatively uniform exposure over long time scales (at 1 GeV over 4 years the exposure varies by approximately 40\% over the entire sky). Combined with a large effective area (approximately 1 m$^2$ between 1 GeV and 1 TeV), this makes it an ideal instrument for detecting a large number of gamma-ray point sources. 
The most complete catalog of point sources is currently the third \Fermi point source catalog  \citep[3FGL, ][]{2015ApJS..218...23A}. 
It contains sources that are significantly detected above 100 MeV and spans the first 4 years of the \Fermi mission. 
We used the 3FGL catalog to search for PBH candidates and to constrain the local PBH evaporation rate.

To find PBH candidates in the 3FGL catalog, we first excluded from further consideration point sources associated with known astrophysical sources (such as blazars). Of 3033 sources in the 3FGL catalog, 1010 are unassociated sources. We also excluded sources that are within $10^\circ$ of the Galactic plane, which removed a further 468 sources. Our analysis was restricted to high-latitude sources because (a) detectable PBHs are expected be distributed isotropically given the detectability distances estimated in Section \ref{sec:sens} while astrophysical sources are concentrated along the Galactic plane, and (b) association of extragalactic sources such as blazars is easier at high latitude 
\citep[see, e.g.][]{2015ApJ...810...14A}.

To test the remaining unassociated 3FGL sources as PBH candidates, we fit their spectra with the time-integrated gamma-ray spectra emitted by a PBH.
For this analysis we used the fluxes of the candidate sources as reported in the 3FGL, which are provided in 5 energy bands (0.1 -- 0.3, 0.3 -- 1, 1 -- 3, 3 -- 10, and 10 -- 100 GeV). The time-integrated PBH spectrum depends on two parameters: initial mass (or temperature) and the distance to the Earth (equivalent to an overall normalization).
We varied these two parameters to obtain the best fit to the PS spectrum in the five energy bands. The quality of the spectral fit to the reported spectrum was determined by calculating the value of the $\chi ^2$ over the five energy bands:

\noindent
\begin{equation}
\chi^2 = \sum_{i = 1}^{5} \frac{(\Phi_i-\Psi_i )^2}{\sigma_i^2}, 
\end{equation}
where $\Phi_i$ is the PBH spectrum integrated over the width of bin $i$ and $\Psi_i$ is the flux in bin $i$ from the 3FGL. Here $\sigma_i$ represents the uncertainty on the flux in bin $i$.
In the 3FGL, flux uncertainty is represented by a 68\% confidence interval; the value of the uncertainty can then be written as the difference between the best-fit flux and either the upper or lower bound.
We set $\sigma_i$ to be the larger of the two in order to be conservative.
We require the value of the best-fit $\chi^2$ to be below the critical value of 11.3, which corresponds to 99\% exclusion for 5 degrees of freedom and 2 parameters. In other words, sources with a $\chi ^2$ value greater than 11.3 have only a 1\% chance of being spectrally consistent with a PBH. After the spectral consistency was computed, 318 sources out of the 542 unassociated candidates remained as PBH candidates.

The candidate sources next underwent a check for proper motion. 
We use the following algorithm to determine the magnitude and significance of proper motion:
\ben
\item
All source-class photons above 1 GeV within $5^\circ$ of the source's reported 3FGL location were collected.
The time range (August 2008 to July 2012) and data reconstruction (\texttt{P7REP\char`_SOURCE\char`_V15}) were consistent with that of the data used to construct the 3FGL.
Since the angular resolution of the \Fermi LAT decreases quickly below 1 GeV, including photons below 1 GeV did not have a significant impact on the final results.
\item 
Data covering a longer time range (August 2008 to July 2017) and a more recent event reconstruction (\texttt{P8R2\char`_SOURCE\char`_V6}) were held in reserve for validation, and was used to test the PBH hypothesis for any sources that passed the proper motion cut.

\item
The expected number of photons $N$ from the source of interest was calculated by multiplying the flux in each energy bin by 
the \Fermi-LAT exposure at the bin's midpoint energy, and summing over the three relevant bins (1 -- 3 GeV, 3 -- 10 GeV, 10 -- 100 GeV). 
\item 
\label{step:prop_motion_likelihood}
In order to estimate the velocity of a PS we compare the maxima of the likelihood function $\La(\vec{x_i},t_i,\vec{x_0},\vec{v_0})$ in two cases: 
fixed $\vec{v_0}=0$ and free $\vec{v_0}$. 
We approximate the point spread function of \Fermi LAT by a Gaussian for simplicity.
The likelihood function is given by multiplying over $N$ photons around the initial position of the source:

\noindent
\begin{equation}
\lb{eq:vlike}
\La = \prod_{i=1}^{N}w_i\times \textup{exp} \big\{\frac{{-(\vec{x_i}-\vec{x_0}-\vec{v_0}t_i)^2}}{{\sigma_i^2}}\big\}
\end{equation}
where $\vec{x_i}$ is the coordinates of the photon, $\vec{v_0}$ is the proper motion of the source, $t_i$ is the photon arrival time, $\vec{x_0}$ is the source location at the beginning of the observation time, and $\sigma_i$ is the 68\% angular containment radius for a photon at energy $E_i$, which is $\sim 0\adeg.7$ at 1 GeV \citep{2013arXiv1304.5456B}.
Here $w_i$ is a weight assigned to each photon, the calculation of which is defined in Section \ref{sec:likelihood}. 

In practice, we use the natural logarithm of the likelihood:
\begin{equation}
\lb{eq:vloglike}
\log \La = \sum_{i=1}^{N}\log w_i-\frac{{(\vec{x_i}-\vec{x_0}-\vec{v_0}t_i)^2}}{{\sigma_i^2}}
\end{equation}

The main difficulty is separating the $N$ photons attributed to the source from background photons.
Our algorithm chooses a 4-dimensional grid of points around an initial value of $\vec{x_0}$ and $\vec{v_0} = 0$, and for each grid point finds the $N$ photons
inside the $5^\circ$ ROI that have the highest contribution to $\log \La$, i.e. the photons which most likely belong to the source given a particular position and velocity.
Therefore the weights $w_i$ do not appear as a prefactor in Eqs \ref{eq:vlike} and \ref{eq:vloglike} because the $N$ best-fit photons change given different assumptions of $\vec{x_0}$ and $\vec{v_0}$.
The best-fit $\vec{x_0}$ and $\vec{v_0}$ are found by maximizing $\Delta \log \La = \log \La - \log \La(\vec{v_0}=0)$ on the grid.
With the additional degrees of freedom from allowing $\vec{v_0}$ to float, the value of $\Delta \log \La$ is always nonnegative. 

We find in MC simulations (described in Section \ref{sec:MClimit}) that this algorithm tends to underestimate the input velocity by $\approx 25\%$;
the best-fit velocity should therefore be considered a lower bound on the true velocity and sufficient for our purpose of separation of moving and stationary sources.
The underestimation occurs because source photons that are far away from the average source position have lower weights and so are less often included in the likelihood calculation.
In their stead are background photons, whose distribution in time is random, and therefore cause the algorithm to favor a slower overall velocity.
\item 
The significance of $\Delta \log \La$ for each source was found by assigning random times $t_i$ drawn from a flat distribution to each photon but fixing the positions $\vec{x_i}$ of all the photons, and reoptimizing $\Delta \log \La$. 
This process was repeated 50 times for each source, and the original value of $\Delta \log \La$ was compared with the distribution of $\Delta \log \La$ for the data sets scrambled in time to find a local significance $\sigma$:

\noindent
\begin{equation}
\sigma = \frac{\Delta \log \La_0 - \overline{\Delta \log \La_s}}{\textup{std}(\Delta \log \La_s)},
\end{equation}
where $\Delta \log \La_0$ is the original value of the improvement in likelihood, $\overline{\Delta \log \La_s}$ is the mean of the scrambled likelihood improvements, and $\textup{std}(\Delta \log \La_s)$ is the standard deviation of the scrambled likelihood improvements.
\item
A cut on the local significance for each source was made at $3.6 \sigma$ which corresponds to a global significance of $2 \sigma$ for 318 sources. 
\een
A single source (3FGL J2310.1$-$0557, see Section \ref{sec:j2310} for a discussion of this source) exceeded this cut on local significance, and the standard deviation of the local significances of the entire set of candidates was 1.03, which is consistent with statistical fluctuations. After examining the data held in reserve for J2310.1$-$0557 (described in Section \ref{sec:j2310}), we concluded that no likely PBH candidates exist in the 3FGL catalog.


\subsection{Calculating Photon Weights}
\lb{sec:likelihood}

The photon weights $w_i$ in equations \ref{eq:vlike} and \ref{eq:vloglike} are defined as the probability that a given photon originated from the candidate PS, and are calculated by performing a standard likelihood optimization with the \Fermi Science Tool {\tt gtlike}\footnote{Science Tools version v10r0p5, available at \url{http://fermi.gsfc.nasa.gov/ssc/data/analysis/software}}.
The model used includes all 3FGL PS within 5$^\circ$ of the candidate source, as well as the standard Pass 7 models for Galactic and isotropic diffuse emission.
The candidate source is modeled as an extended source with a radial Gaussian profile with $\sigma=0.25^\circ$ instead of a PS, in order to account for the possibility of proper motion.
The data were binned into three logarithmically spaced energy bands between 1 GeV and 100 GeV and in $0.1^\circ \times 0.1^\circ$ spatial pixels.
After the model was optimized by {\tt gtlike}, weights were assigned to each photon (described its coordinates $x$, $y$, and energy $E$) by calculating the fraction of the flux belonging to the candidate source in each pixel:

\begin{equation}
w_{x,y,E} = \frac{\Phi'(x,y,E)}{\sum_i \Phi_i(x,y,E)}
\end{equation}
where $\Phi'(x,y,E)$ is the predicted flux from the candidate source in the pixel and $\Phi_i(x,y,E)$ are the fluxes from all the sources in the model.
In addition, the 3FGL PS were masked by assigning a weight of zero to all the photons which fell in a pixel more than 1$^\circ$ from the candidate source position where the summed contribution of the non-candidate PS fluxes exceeded 10\% of the total flux in that pixel. This meant that all the photons in the calculation had a high probability of originating either from the candidate source or the diffuse background.

The weighting has little impact on the reconstruction of proper motion because individual photon weights do not change as the likelihood maximization from step \ref{step:prop_motion_likelihood} optimizes $\vec{x_0}$ and $\vec{v_0}$. However, weighting the photons in this way prevents the algorithm from interpreting photons from nearby sources as originating from the candidate source. Without weighting, we found that flaring nearby sources could mimic a moving source and therefore lead to false positive results.

\subsection{J2310.1$-$0557}
\lb{sec:j2310}
The source J2310.1$-$0557 passed the proper motion cut with a significance of 4.2$\sigma$, and was therefore investigated further.
Approximately 9 years (August 2008 to July 2017) of Pass 8 (\texttt{P8\char`_SOURCE\char`_V6}) data above 1 GeV in an ROI of 5$^\circ$ around the source location were collected.
The increased statistics and improved angular resolution of the Pass 8 data set clearly indicated that J2310.1$-$0557 lies approximately 1$^\circ$ away from a separate, highly variable source of gamma rays which is not in the 3FGL catalog.
This source flared brightly (approximately 150 photons) on 2011 March 7, near the end of the 3FGL time period but was quiet for the remainder of the period.
We found that the position of the source was consistent with the Sun, which flared brightly on the same date \citep{2011ATel.3214....1A}.
gamma-ray emission from the Sun and Moon not included in our models of the ROI.
The effect of the solar flare near a candidate PBH was to mimic a moving source, which explains why the proper motion algorithm returned a positive result. 
Because the sources in the Monte Carlo simulation described in Section \ref{sec:MClimit} are placed at random points on the sky, we expect that similar false positives will occur in the simulations.
Therefore, we report the upper limit on PBH evaporation rate as if one source passed our criteria, even though J2310.1$-$0557 is not a good PBH candidate. Incidentally after the publication of the 3FGL source list, J2310.1$-$0557 was found to be a millisecond pulsar\footnote{See https://confluence.slac.stanford.edu/display/SCIGRPS/LAT+Pulsations+from+PSR+J2310-0555}.

%% file: 4PBH_MClimits.tex
\section{\Fermi-LAT limits on PBH\MakeLowercase{s}}
\label{sec:MClimit}

We used Monte Carlo (MC) simulations to derive the efficiency for detecting PBHs, and used the efficiency to place upper limits on the local PBH evaporation rate.
We generated a sample of PBHs within 0.08 pc of the Earth with uniform spatial density and random velocities with an average speed of
$250\rm\,km\, s^{-1}$, which is close to an upper bound on orbital velocity of the Sun around the Galactic center
\citep{1999MNRAS.310..645W, 2008ApJ...684.1143X, 2009PASJ...61..227S, 2010ApJ...720L.108G, 2010MNRAS.402..934M, 2011MNRAS.414.2446M},
and 3-dimensional velocity dispersion equal to the local velocity dispersion of dark matter, $270\rm\,km\, s^{-1}$ \citep{2010JCAP...02..030K}.
At the end of this section, we also derive the limits for different assumptions about the PBH distribution, such as the relative velocity and 
velocity dispersion, to estimate the corresponding systematic uncertainty.

We assume the PBH population has a constant rate of PBH evaporations,
$\dot{\rho}_{\rm PBH} = const.$
We also assume a uniform PBH density distribution in the vicinity of the Earth. 
A constant rate of evaporation implies that the derivative of the PBH density is related to the PBH temperature as:

\noindent
\be
\lb{eq:Tdistr}
\frac{d \rho_{\rm PBH}}{d T} \propto T^{-4}.
\ee



\begin{figure}[htbp]
\begin{center}
\epsfig{figure = 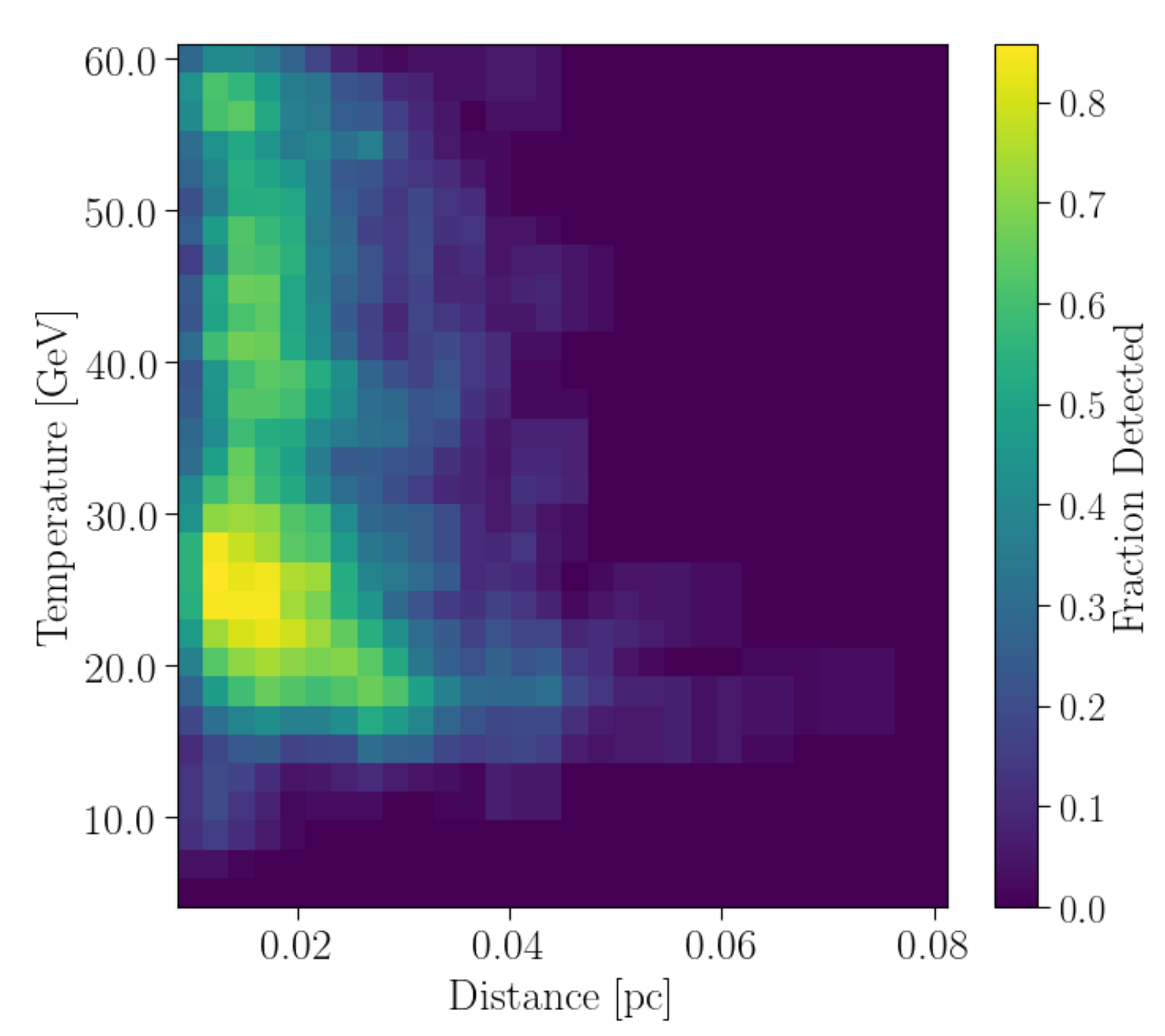, scale=\onepic}
\noindent
\end{center}
\caption{\small 
\label{fig:det_map}
Fraction of simulated PBHs which are detected as a point source with a spectrum compatible with a PBH evaporation spectrum and with significant proper motion. The detectability peaks for PBHs with initial temperatures above 16.4 GeV because the lifetime of a 16.4 GeV PBH is 4 years, which is the same as the observation period of the 3FGL. Few PBHs are detected past a distance of 0.05 pc or below 10 GeV.}

\end{figure}

The following steps were performed in the derivation of the PBH evaporation rate limit:
\vspace{-2mm}
\ben
\item
\label{item:make_pbhs}
A sample of PBHs $(T_i, \vec{x}_i, \vec{v}_i)$ was simulated with temperatures $T_i>5\: {\rm GeV}$ and $T_i<60\: {\rm GeV}$ distributed according to Equation (\ref{eq:Tdistr}), and distances $R_i$ within $R < 0.08$ pc around the Earth. 
The velocities $v_i$ of the sample PBHs were distributed with mean equal to the orbital velocity of the Sun, 
$v_{\rm rot} = 250\; {\rm km\: s^{-1}}$,
and dispersion $v_{\rm disp} = 270\; {\rm km\: s^{-1}}$.
\item
For each PBH, we simulated the detection of the photons emitted over the 4 year 3FGL time period, consistent with the PBH evolution. The energies were distributed according to the instantaneous PBH spectrum of the appropriate temperature, and the positions of the photons were smeared according to the 
\Fermi-LAT point-spread function (modeled as a Gaussian distribution).
The emission spectra of PBHs $\Phi(E,t)$ are discussed in Appendix \ref{app:PBH_spectrum}. 
We used a time step of $\Delta t = 1$ day in modeling the evolution of the PBH position and temperature, 
and the number of photons detected by the \Fermi LAT each day was given by a Poisson random value with a mean of 

\noindent
\begin{equation}
\overline{N(t)} = \frac{ \Delta t}{4\pi R^2}\int_{\textup{E=100 MeV}}^{\textup{E=500 GeV}} {\Phi(E,t)} A(E) dE,
\end{equation}
where $A$ is defined as the average \Fermi-LAT exposure per unit time at the position of the simulated PBH, and $R$ is the distance from the Earth.
The energy of each photon was found by random sampling of $\Phi(E,t)\times A(E)$. 
\Fermi LAT has relatively uniform exposure on time periods longer than 1 day.
\item
\label{item:detection}
The list of simulated PBH photons was concatenated to the real photons present within $5^\circ$ of the final location of the PBH, with the same data selection as the 3FGL. A likelihood fit using the \Fermi Science Tool {\tt gtlike} was performed in a $7^{\circ}\times7^{\circ}$ ROI centered at the same location, using a model of the sky which included all 3FGL sources within $5^\circ$ of the ROI center, as well as models of the isotropic diffuse and Galactic diffuse emission\footnote{The models used were the standard Pass 7 (for consistency with the 3FGL) diffuse emission models available from \url{https://fermi.gsfc.nasa.gov/ssc/data/access/lat/BackgroundModels.html}}. The PBH was modeled as a source with a LogParabola spectrum, with fitting parameters restricted to the ranges $1.2<\alpha<3.0$ and $0.0<\beta<1.0$. Once the likelihood maximization was complete, the PBH was considered detected if its TS value was greater than 25, which is consistent with the 3FGL cutoff.
\item
\label{item:spectrum}
If the PBH source was detected, the results from the likelihood fit were used to find the source flux in the five energy bins reported in the 3FGL catalog. The spectral consistency with a PBH spectrum was then calculated in the same way as described in Section \ref{sec:search}.
\item 
\label{item:motion}
If the source was found to be spectrally consistent with a PBH, the significance of any proper motion was evaluated by the algorithm described in Section \ref{sec:search}. The combined efficiency of steps \ref{item:detection}$-$\ref{item:motion} is displayed in Figure \ref{fig:det_map}. We smoothed the results by convolving the detectability map with a 3$\times$3 matrix of ones, which had a minor ($\approx$ 8\%) impact on the resulting limit. The impact of fluctuations was quantified by observing the change in the limit as the number of simulations increased; we found that an increase of the number of simulations by 100\% had less than a 20\% change in the resulting limit.
\item 

\lb{step:eff}
To derive an upper limit on the number of PBH evaporations in our search region, we begin with the number of expected detections:

\noindent
\begin{equation}
\label{eq:N_det}
N = \rho \epsilon V,
\end{equation}
where $\rho$ is the true density of PBHs and $V$ is the volume searched.
$\epsilon$ is the average PBH detection efficiency in time $t = 4$ yr and within the search volume $V$ (a sphere with radius 0.08 pc, with the wedge corresponding to $|b|<10^\circ$ removed); 
it is calculated by taking the mean over the pixels in Figure \ref{fig:det_map} with the weight $R^2 T^{-4}$:

\noindent
\begin{equation}
\epsilon = \frac{\iint \epsilon (R, T) \frac{R^2}{T^4} \,dR\,dT }{\iint \frac{R^2}{T^4}\,dR\,dT },
\end{equation}
where the integrals run over the space of parameters described in step \ref{item:make_pbhs}. 
Equation \ref{eq:N_det} can be inverted to find the PBH density $\rho$ as a function of the number of detections $N$, or the upper limit on $\rho$ given an upper limit on $N$. Given that one PBH candidate passed the selection criteria described in Section \ref{sec:search}, we set an upper limit $N < 6.64$, 
which is the 99\% confidence upper limit on the mean of a Poisson distribution with 1 observed event.

\item
\label{item:limit}
We convert the upper limit on $\rho$ to an upper limit on $\dot{\rho}$ by finding the fraction $f$ of PBHs that would have evaporated during the search time 
$t$. 
Given a time of observation of 4 years, we find that all PBHs with initial temperature above 16.4 GeV would evaporate. Therefore,

\noindent
\begin{equation}
f = \frac{\int_{\textup{16.4 GeV}}^{\textup{60 GeV}} T^{-4} \,dT}{\int_{\textup{5 GeV}}^{\textup{60 GeV}} T^{-4} \,dT}.
\end{equation}
We calculate the 99\% upper limit on $\dot{\rho}_{\rm PBH}$ to be:

\noindent
\begin{equation}
\dot{\rho}_{\rm PBH} < f\frac{6.64}{\epsilon V t} = 7.2 \times 10^3 \textup{ pc}^{-3} \textup{ year}^{-1}.
\end{equation}

\item
We estimated the systematic uncertainties arising from the uncertaintes in the PBH spectrum by varying the overall normalization of the PBH spectrum (see Appendix \ref{app:PBH_spectrum}) and varying the velocity distributions of the Milky Way disk and DM halo. We consider two scenarios (``aggressive" and ``conservative") which give the best and worst sensitivity, respectively. Steps \ref{item:make_pbhs}$-$\ref{item:limit} are then repeated to find the resulting limit. The parameters of the aggressive and conservative models, as well as the resulting limits, are listed in Table \ref{table:systematics}.
\begin{table}[h]
\begin{center}
\begin{tabular}{|c c c c c|}
\hline
Model & Spectrum Normalization & Orbital Velocity (${\rm km\: s^{-1}}$) & DM Halo Velocity (${\rm km\: s^{-1}}$) & Limit\\
\hline
Aggressive & $\frac{0.45}{0.35}$ & 100 & 150 & $4.8\times 10^3  \textup{ pc}^{-3}  \textup{yr}^{-1}$ \\
Conservative & $\frac{0.25}{0.35}$ & 300 & 350 & $15.3\times10^3 \textup{ pc}^{-3} \textup{ yr}^{-1}$ \\
\hline
\end{tabular}
\caption{Parameters used in estimation of systematic uncertainty. 
To be more conservative in the estimates of the systematic uncertainties,
we have tested the ranges of orbital velocities and the DM dispersion velocities which are larger than most of the values reported
in the literature
\citep{1999MNRAS.310..645W, 2008ApJ...684.1143X, 2009PASJ...61..227S, 2010ApJ...720L.108G, 2010JCAP...02..030K,
2010MNRAS.402..934M, 2011MNRAS.414.2446M}.}
\label{table:systematics}
\end{center}
\end{table}

The limit including the systematic uncertainties is

\noindent
\be
\lb{eq:ev_rate}
\dot{\rho}_{\rm PBH} < (7.2^{+8.1}_{-2.4} ) \times 10^3 \textup{ pc}^{-3} \textup{ yr}^{-1}.
\ee
\een

%% file: 5PBH_discussion.tex
\section{Discussion and Conclusions}
\lb{sec:discuss}

The potential existence of PBHs 
that emit detectable Hawking radiation is one of the most intriguing 
features of some theories of cosmological evolution.
In addition to providing evidence for these theories, the possibility for direct observation of Hawking radiation (which would be a major discovery in its own right) was our main motivation for the search of PBHs with the \Fermi LAT.

In this paper we searched for individual PBHs among 3FGL catalog sources.
We performed calculations showing that the characteristic distance to a detectable PBH is of order $\sim 0.03$ pc,
in which case the proper motion of the PBH must be taken into account.
We developed a new algorithm which can detect the proper motion of a point source in the presence of a known background.
We found several 3FGL sources that have spectra consistent with PBHs,
but none of these sources exhibit proper motion which would be the smoking-gun signature of a PBH in the \Fermi-LAT sensitivity domain.
As a result, we derived upper limits on the local PBH evaporation rate.

To derive the efficiency of a PBH passing our selection criteria, we developed a Monte Carlo that simulates PBHs with realistic velocity distribution and initial temperature distribution expected for the steady-state evaporation rate of the PBHs. The efficiency was then used to find the upper limits on the local PBH evaporation rate.
The inferred systematic uncertainties are related to the uncertainties in the PBH gamma-ray spectrum as well as the uncertainties in the Galactic rotational velocity and the DM velocity dispersion. We calculated upper limits in scenarios for which these parameters covered a wide range of reasonable values.

\begin{figure}[htbp]
\begin{center}
\epsfig{figure = 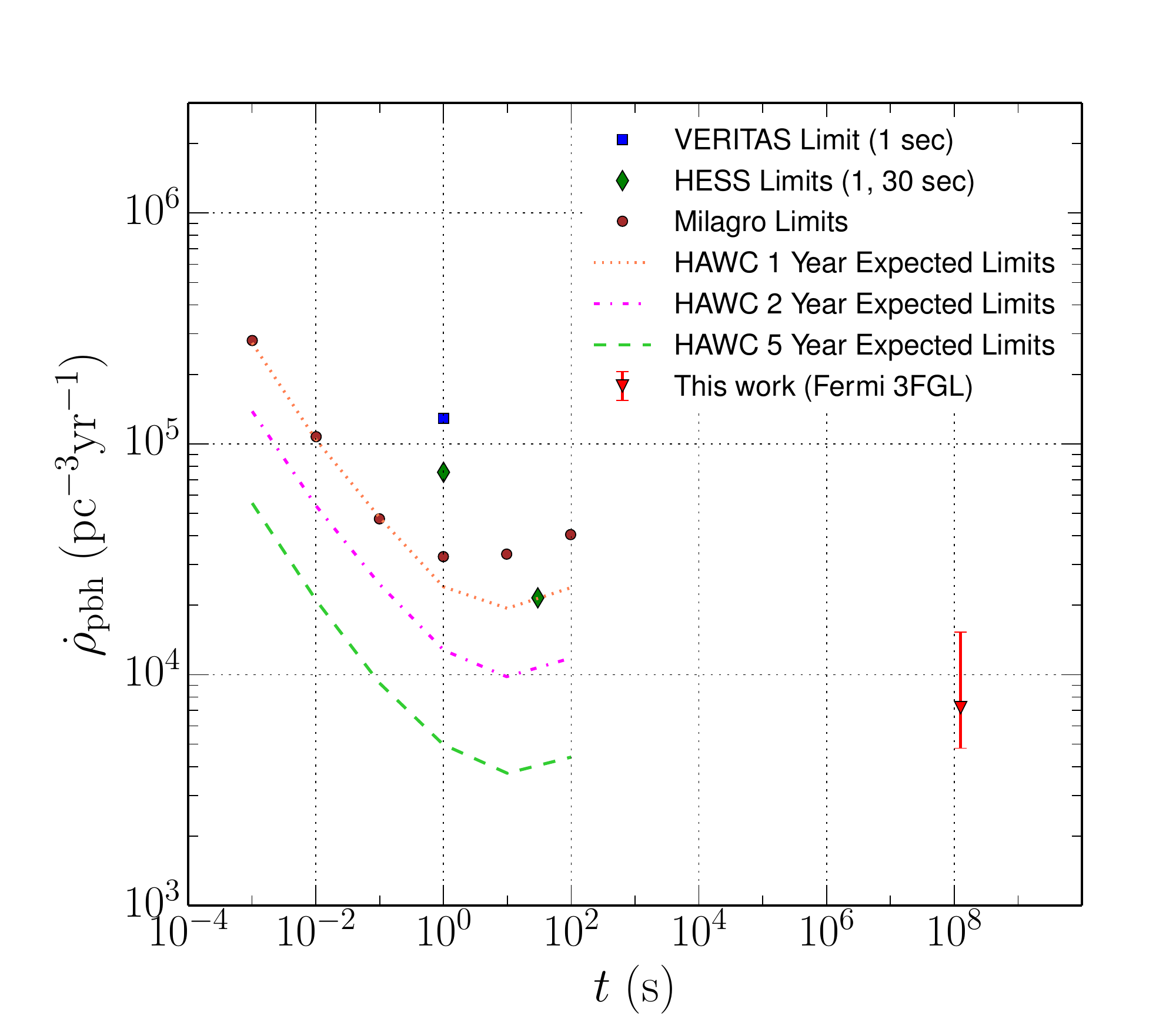, scale=\onepic}
\noindent
\caption{\small 
\label{fig:limit_comparison}
Comparison of the \Fermi-LAT 99\% confidence upper limit with the limits from 
VERITAS \citep{2012JPhCS.375e2024T},
H.E.S.S. \citep{2013arXiv1307.4898G},
Milagro \citep{2015APh....64....4A},
and the expected limits from HAWC \citep{2015APh....64....4A}.
The error bars around the \Fermi-LAT limit correspond to the systematic uncertainty described in the text. 
}
\end{center}
\end{figure}

In Figure \ref{fig:limit_comparison}, we compare the \Fermi-LAT upper limit 
on the rate of PBH evaporations with the limits from Cherenkov telescopes and observe that they are similar. 
Although ground-based gamma-ray observatories are sensitive to timescales of a minute or less, and \Fermi LAT is sensitive to timescales of months to years, both the Cherenkov telescopes and the \Fermi LAT are probing the same quasi-stationary population of PBHs.

The local evaporation rate can be related to the local mass density (relative to the critical density $\rhoc$) as
\citep{1991Natur.353..807H} 

\noindent
\be 
\Ombh^{\rm loc} = \frac{M_*}{(\alpha - 2)}
\frac{\tau_0}{\rhoc} \rbh, 
\ee 
where $\alpha$ is the index of the initial distribution of PBH masses, ${dn}/{dM} \sim M^{-\alpha}$,
and  $M_*$ is the mass of a PBH with the lifetime equal to the age of the Universe.
In the following, we will assume $\alpha = 2.5$, which corresponds to PBHs formed in the 
radiation-dominated era \citep{1991Natur.353..807H}.
The local density of PBHs corresponding to the evaporation rate in Equation (\ref{eq:ev_rate}) is
$\rm\Ombh^{loc} \leq (3.6^{+4.1}_{-1.2}) \times 10^{2}$.
In general, the distribution of the PBH masses does not need to follow a power-law function.
If there is a short period of low-pressure dustlike equation of state or a phase transition (see, e.g., the discussion in \cite{2010PhRvD..81j4019C} and references therein), 
then the PBH masses will be distributed around the mass in Equation \ref{eq:bh_mass}, 
where $t$ is the time of the PBH formation.
In order for the PBHs to be evaporating now we need $M \sim 10^{15}$ g which corresponds to the formation
time  $t \sim 10^{-23}$ s after the Big Bang \citep{2010PhRvD..81j4019C}.
Thus, in general, limits on local PBH evaporation can constrain models where PBHs are formed with lifetimes close to the lifetime of the Universe, i.e., the PBHs which are close to evaporation now.
This requires certain ``fine-tuning" of the time when the PBHs are formed.
If, for example, the PBHs are formed close to the QCD phase transition with $t_Q \sim 10^{-4}$ s and $T_Q \sim 100$ MeV,
then the PBH masses are $M \gtrsim 5 M_\odot \approx 10^{34}$ g \citep{2014PhRvD..89b1301D}.
These PBHs have a temperature much smaller than $10^{-7}$ K and lifetime much longer than the lifetime of the Universe,
i.e., they cannot be detected by the \Fermi LAT.

The local density of PBHs is expected to be enhanced
compared to the average density in the Universe in a similar way that
the density of DM in the Galaxy is larger than the average
density of DM in the Universe.  The enhancement factor for DM near the Sun \citep{2012ApJ...756...89B} 
compared to the average DM density \citep{2013ApJS..208...19H} 
is $k \sim 2.2 \times 10^5$.
With this enhancement, 
the limit on the average PBH density is
$\Ombh \leq \rm (1.5^{+1.7}_{-0.5}) \times 10^{-3}$.
This limit is several orders of magnitude less constraining than the limits obtained from extragalactic and Galactic gamma-ray backgrounds
$\Ombh \leq 10^{-8} - 5 \times 10^{-10}$ \citep{1998PhR...307..141C, 2009A&A...502...37L, 2010PhRvD..81j4019C}.
The latter limits are calculated either by integrating the PBH evaporations inside the visible Universe or inside the halo of our Galaxy,
i.e., they are derived on kpc to Gpc scales, while the limit in this paper is derived for distances less than a fraction of a pc.


The limit on the average current density of PBHs
can be translated to a limit on the density of PBHs at the time of
formation, $\beta \sim 10^{-18} \Ombh\sqrt{M / 10^{15}{\rm g}}$
\citep{2005astro.ph.11743C, 2010PhRvD..81j4019C}.  
This limit, in turn, can be used to constrain
the spectrum of density fluctuations in the early Universe
\citep{2010PhRvD..82d7303J, 2013PhRvD..87j3506L, 2013AIPC.1548..238T}.  
In some cases, non-observation of PBHs provides the only 
way to limit theories of inflation, especially the theories
that predict large fluctuations of density at small distances
\citep[e.g.,][]{2013PhRvD..87j3506L}.

%% file: acknowledgement.tex
\acknowledgments

The \Fermi-LAT Collaboration acknowledges generous ongoing support
from a number of agencies and institutes that have supported both the
development and the operation of the LAT as well as scientific data analysis.
These include the National Aeronautics and Space Administration and the
Department of Energy in the United States, the Commissariat \`a l'Energie Atomique
and the Centre National de la Recherche Scientifique / Institut National de Physique
Nucl\'eaire et de Physique des Particules in France, the Agenzia Spaziale Italiana
and the Istituto Nazionale di Fisica Nucleare in Italy, the Ministry of Education,
Culture, Sports, Science and Technology (MEXT), High Energy Accelerator Research
Organization (KEK) and Japan Aerospace Exploration Agency (JAXA) in Japan, and
the K.~A.~Wallenberg Foundation, the Swedish Research Council and the
Swedish National Space Board in Sweden.
 
Additional support for science analysis during the operations phase is gratefully
acknowledged from the Istituto Nazionale di Astrofisica in Italy and the Centre
National d'\'Etudes Spatiales in France. This work performed in part under DOE
Contract DE-AC02-76SF00515.

The authors would like to thank Jane MacGibbon for numerous helpful discussions.




\facility{\textit{Fermi}}.

\software{Astropy \citep[\url{http://www.astropy.org},][]{2013A&A...558A..33A}, 
matplotlib \citep{Hunter:2007}}.

%% file: App_PBH_spectra.tex
\section{Spectrum of Gamma Rays from Primordial Black Holes}
\label{app:PBH_spectrum}

One of the most important factors in the systematic uncertainty of the limit on PBH evaporation is the spectrum
of emitted gamma rays.
The spectrum of fundamental particles emitted by the black hole was computed by \cite{1975CMaPh..43..199H}.
Quarks and gluons emitted by PBHs hadronize into mesons and baryons, which subsequently decay into stable particles.
The spectra of the stable particles emitted by PBHs at high temperatures were computed by \cite{1990PhRvD..41.3052M}.

\begin{figure}[htbp] 
\begin{center}
\epsfig{figure = 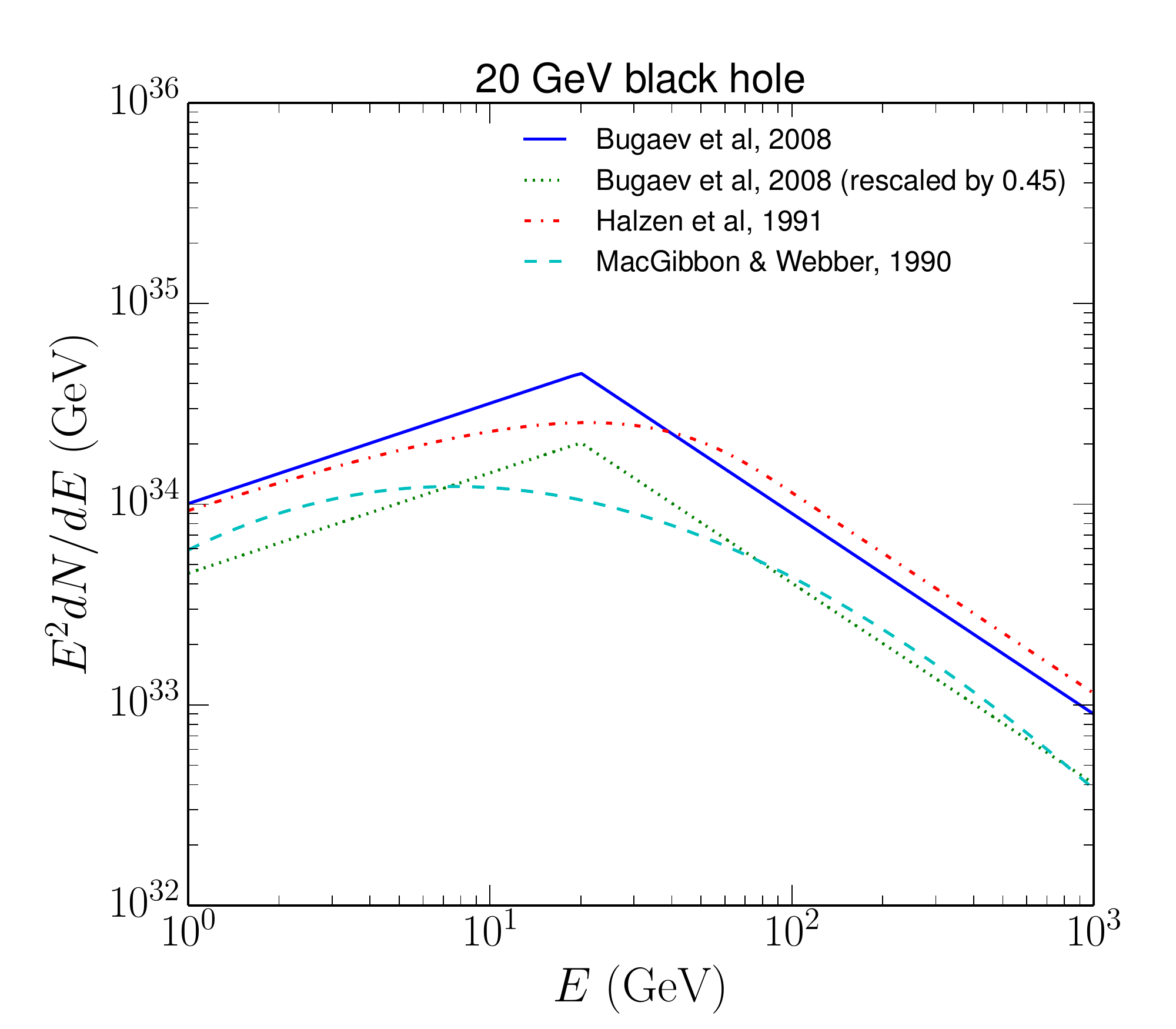, scale=\twopic}
\epsfig{figure = 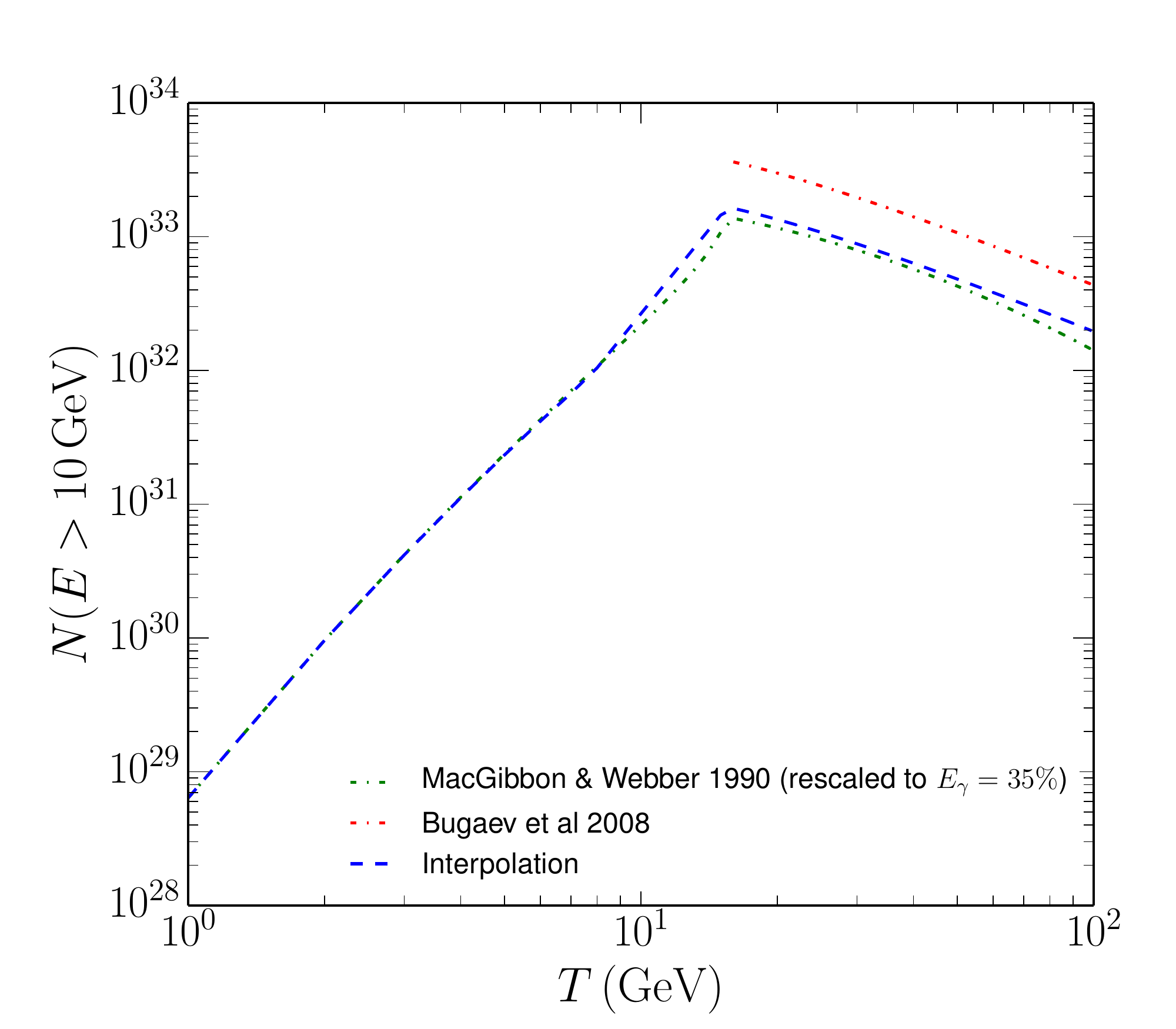, scale=\twopic}
\noindent
\caption{\small 
\label{fig:PBH_spectra}
Left: spectrum of a PBH with initial temperature $T = 20$ GeV integrated over its lifetime of 2.1 years.
Right: total number of photons emitted above 10 GeV during 4 years as a function of initial PBH temperature.
See text for the description of the ``Interpolation" curve.
}
\end{center}
\end{figure}

To obtain the spectra of gamma-ray emission from PBHs we use the values given in \cite{1990PhRvD..41.3052M}
for  $T_{\rm PBH} = 0.3, 1, 10, 50, 100$ GeV and interpolate for different values of PBH temperatures.
To cross-check numerical calculations, we use two different interpolation methods: 
\bi
\item
We use an analytic approximation by fitting the PBH emission rate $\dot{N}_\gamma (E, T)$ with a cubic log polynomial 

\noindent
$$\log \dot{N} = c_0(T) + c_1(T) \log x + c_2(T) (\log x)^2 + c_3(T) (\log x)^3,$$
where $x = E / T$ and interpolate the fit coefficients $c_i(T)$
(we use this appriximation in Section \ref{sec:sens}).
\item
We create a table for a set of $E$ and $T$ values and use a 2-dimensional interpolation directly from the results in \cite{1990PhRvD..41.3052M}
(this approximation is used  in Sections \ref{sec:search} and \ref{sec:MClimit}).
\ei
We compare the first interpolation with other parametrizations available in the literature \citep{1991Natur.353..807H, 2013arXiv1308.4912U}
in Figure \ref{fig:PBH_spectra} left.
There seems to be a rather significant discrepancy in the total energy emitted in gamma rays.
For instance, \cite{1990PhRvD..41.3052M} found that 24 -- 25\% of the energy is emitted in gamma rays for a large range of
initial temperatures, while analytical integration of the parametrization \citep{2008ICRC....3.1123B, 2008AstL...34..509P}

\noindent
\be
\frac{dN_{\gamma}}{dE_{\gamma}} = 9 \times 10^{35}
\left\{
\ba{ll}
\left( \frac{\rm 1\, GeV}{T_{\tau}} \right)^{3/2} \left( \frac{\rm 1\, GeV}{E_{\gamma}} \right)^{3/2} {\rm GeV}^{-1} &
E_{\gamma} <  T_{\tau}; \\
\left( \frac{\rm 1\, GeV}{E_\g} \right)^{3} {\rm GeV}^{-1} &
E_{\gamma} \geq T_{\tau}
\ea
\right.
\ee
gives about 47\% of the energy in gamma rays.
A more-recent parametrization of the gamma-ray spectra at temperatures $\gtrsim 1$ TeV 
\citep[Figure 11 of][]{2016APh....80...90U}
gives about 35\% of the energy in gamma rays.

For our baseline model, we take the spectrum of \cite{1990PhRvD..41.3052M} rescaled to give 35\% of energy
in gamma rays to match the more-recent calculation in \cite{2016APh....80...90U}.
For the calculations, we use the interpolation presented in Figure \ref{fig:PBH_spectra} (right).
At low temperatures, the ``Interpolation" curve is calculated by taking instantaneous PBH spectrum from \cite{1990PhRvD..41.3052M}
rescaled to 35\% energy going to gamma rays times 4 years.
At temperatures above the temperature of a PBH with the 4 years lifetime,
the ``Interpolation" curve is the \cite{2008ICRC....3.1123B} parametrization rescaled by 0.45.
There is a good agreement between the integrated spectra of \cite{1990PhRvD..41.3052M} (rescaled to 35\%  efficiency)
and the ``Interpolation" curve.

